\documentclass[11pt]{article}
\usepackage{amsthm, mathrsfs,amssymb,clrscode,amsmath}
\usepackage{quantph}  
\newcommand{\abs}[1]{\vert #1 \vert}
\newcommand{\gen}[1]{\langle #1 \rangle}

\newcommand{\Integers}{\mathbb Z}
\newcommand{\Int}{\mathbb Z}

\newcommand{\bound}[2]{\in\{ #1,\ldots,#2\}}
\newcommand{\Sp}{\mathscr{S}}

\newcommand{\Dd}{\mathscr{D}}

\newcommand{\st}{\:\vert\:}
\newcommand{\triv}{\{e\}}
\newtheorem{theorem}{Theorem}[section]
\newtheorem{proposition}{Proposition}[section]
\newtheorem{definition}{Definition}[section]
\newtheorem{lemma}{Lemma}[section]
\begin{document}
\begin{center}

{\LARGE \bf Efficient Isomorphism Testing for \\a Class of  Group Extensions}\vspace{4mm}

{\Large Fran{\c c}ois Le Gall}\vspace{3mm}

{\it ERATO-SORST Quantum Computation and Information Project}\\
{\it Japan Science and Technology Agency, Tokyo}\\


email: legall@qci.jst.go.jp\vspace{1mm}

\setlength{\baselineskip}{11pt}
      \begin{quotation}
\noindent{\bf Abstract.}\hbox to 0.5\parindent{}
The group isomorphism problem asks whether two given groups are isomorphic or not.
Whereas the case where both groups are abelian is well understood and can be solved 
efficiently, very little is known about the complexity of isomorphism testing for nonabelian 
groups. In this paper we study this problem for a class of groups corresponding to one of the
simplest ways of constructing nonabelian groups from abelian groups: the groups that are 
 extensions of an abelian group $A$ by a cyclic group 
$\Int_m$. 
We present an efficient algorithm solving the group isomorphism problem for all the groups 
of this class such that the order of $A$ is coprime with $m$. 
More precisely, our algorithm runs in time almost linear in the orders of the input groups and
works in the general setting where the groups are given as black-boxes. 
\end{quotation}
\setlength{\baselineskip}{11pt}
      \begin{quotation}
\end{quotation}
\end{center} 

\section{Introduction}
The group isomorphism problem is the problem of deciding, for two given groups
$G$ and $H$,  whether there exists an isomorphism between $G$ and
$H$, i.e.~a one-one map preserving the group operation. 
This is a fundamental problem in computational group theory but
little is known about its complexity. 
It is known that the group isomorphism problem (for groups given by their multiplication tables) reduces to the graph isomorphism problem \cite{Kobler+93}, 
and thus the group isomorphism problem is in the complexity class $NP\cap coAM$  (since the graph isomorphism problem is in this class  \cite{BabaiSTOC85}).
Miller \cite{MillerSTOC78} has developed a general technique to check group isomorphism in time $O(n^{\log n+O(1)})$, where $n$ denotes the size of the input groups and Lipton, 
Snyder and Zalcstein \cite{Lipton+76} have given an algorithm working in $O(\log^2{n})$ space. However, no polynomial algorithm is known for the general case of this problem.

Another line of research is the design of algorithms solving the group isomorphism problem for particular classes of groups.
For abelian groups polynomial time algorithms follow directly from efficient algorithms for the computation of Smith normal form of integer 
matrices  \cite{Kannan+SICOMP79,Chou+SICOMP82}. More efficient methods have been given by Vikas \cite{Vikas96} and 
Kavitha \cite{KavithaJCSS07} for groups given by their multiplication tables.
The current fastest algorithm solving the abelian group isomorphism problem for groups given as black-boxes has been 
developed by Buchmann and Schmidt \cite{Buchmann+05} and works in time $O(n^{1/2}(\log n)^{O(1)})$.
However, as far as nonabelian groups are concerned, very little is known.
For solvable groups Arvind and Tor{\'a}n \cite{Arvind+CCC04} have shown that the group isomorphism problem is in $NP\cap coNP$ 
under certain complexity assumptions
but, to our knowledge, the only polynomial-time algorithm testing isomorphism of a nontrivial class of nonabelian groups
is a result by Garzon and Zalcstein \cite{Garzon+JCSS91}, and holds for a very restricted class.

In this work we focus on the complexity of the group isomorphism problem over classes of 
nonabelian groups. Since for abelian groups the problem can be solved efficiently, we study one of the most natural 
next targets: cyclic extensions of abelian groups. 
Loosely speaking such extensions are constructed by taking an abelian group $A$ and adding one element $y$ that, in general, does not commute with the elements
in $A$. 
More formally the class of groups we consider in this paper, denoted $\Sp$, is the following.
\begin{definition}
Let $G$ be a finite group. 
We say that $G$ is in the class $\Sp$ if there exists a normal abelian subgroup $A$ in $G$ and 
an element $y\in G$ of order coprime with $\abs{A}$ such that $G=\gen{A,y}$.
\end{definition}
In technical words $G$ is an extension of an abelian group $A$ by a  cyclic group $\Int_{m}$ with 
$gcd(\abs{A},m)=1$. We will say more about mathematical properties of these extensions in Section \ref{section_prelim}. 
For now, we mention that this class of groups includes all the abelian groups and many non-abelian groups too.  
For example, for $A=\Int_3^4$ and $m=4$ we will show that there are exactly $9$ isomorphism classes in $\Sp$.

A group can be represented on a computer in different ways. In this paper we use the black-box setting introduced by Babai and
Szemer\'edi \cite{Babai+FOCS84}, which
is one of the most general models for handling groups,  and particularly convenient to discuss algorithms running in
sublinear time. 
In order to state precisely the running time of our algorithm, we introduce the following definition.
\begin{definition}
For any group $G$ in the class $\Sp$, let $\gamma(G)$ be the smallest integer $m$ such that  $G$ is 
an extension of an abelian group $A$ by the  cyclic group $\Int_{m}$ with $gcd(\abs{A},m)=1$.
\end{definition}
The main result of this paper is the following theorem. 
\begin{theorem}\label{theorem_main}
There exists a deterministic algorithm checking whether two groups $G$ and $H$ in the class $\Sp$ given as black-box groups
are isomorphic and, if this is the case, computing an isomorphism from $G$ to $H$.  Its running time has for upper bound $(\sqrt{n}+\gamma)^{1+o(1)}$, where $n=min(\abs{G},\abs{H})$ and 
$\gamma=min(\gamma(G),\gamma(H))$.
\end{theorem}

Notice that, for any group $G$ in the class $\Sp$, the relation $\gamma(G)\le \abs{G}$ holds.
Then the complexity of our algorithm has for upper bound $n^{1+o(1)}$, and is almost linear in the
size of the groups.
Another observation is that, if $\gamma=O(n^{1/2})$, then the complexity of our algorithm is $n^{1/2+o(1)}$
and is of the same order as the best known algorithm testing isomorphism of abelian groups \cite{Buchmann+05} in the black-box setting.
This case $\gamma=O(n^{1/2})$ corresponds to the rather natural problem of testing isomorphism of extensions of a large 
abelian group by a small cyclic group.

The outline of our algorithm is as follows. 
Since a group $G$ in the class $\Sp$ may in general be written as the extension of 
an abelian group $A_1$ by a cyclic group $\Int_{m_1}$ and as the extension of 
an abelian group $A_2$ by a cyclic group $\Int_{m_2}$ with 
$A_1\not\cong A_2$ and $m_1\neq m_2$, we introduce (in Section \ref{section_standard}) the concept of a standard
decomposition of $G$, which is an invariant for the groups in the class $\Sp$ in the sense that
two isomorphic groups have similar standard decompositions (but the converse is false). 
We also show how to compute a standard decomposition of $G$ efficiently.
This allows us to consider only the case where $H$ and $G$ are two extensions of the same abelian group $A$ by the same
cyclic group $\Int_m$.
One of the main technical contributions of this paper is an efficient algorithm 
that tests whether two automorphisms of order $m$ in the automorphism group of $A$ are conjugate or not
(Section \ref{section_conjugacy}).
Finally, we present a time-efficient reduction from the problem of 
testing whether $G$ and $H$ are isomorphic to an instance of the above conjugacy problem (Section \ref{section_algorithm}). 

\noindent{\bf Remark.}
The problem of deciding whether two group extensions are isomorphic has been studied 
by mathematicians for a long time. Mathematical results on the isomorphism of semidirect products are known, 
e.g.~\cite{Kunzennyi74}, but to our knowledge do not give computationally efficient criterions for the isomorphism 
of groups in the class $\Sp$. 
More generally several algorithms for the group isomorphism problem performing relatively well 
in practice are known and have been implemented in computational group theory softwares (GAP, MAGMA,...). 
The main works in this area are the algorithms developed by Smith for solvable groups \cite{Smith94} 
and  by O'Brien \cite{ObrienJSymb94} for $p$-groups. However these 
algorithms involve computation in groups of size exponential in $n$, e.g.~the automorphism groups or 
the cohomology groups, and no rigorous analysis of their time complexity is available.

\section{Preliminaries}\label{section_prelim}
We assume that the reader is familiar with the basic notions of group theory and state without proofs 
basic definitions and properties of groups we will use in this paper. 

Let $G$ be a finite group (in this paper we will consider only finite groups).
For  any subgroup $H$ and any normal subgroup $K$ of $G$ we denote by $HK$ the subgroup $\{hk\st  h\in H,k\in K\}=\{kh\st  h\in H,k\in K\}$. 
Given a set $S$ of elements of $G$, the subgroup generated by the elements of $G$ is written $\gen{S}$.
We say that two elements $g_1$ and $g_2$ of $G$ are conjugate if there exists an element $y\in G$ such that
$g_2=yg_1y^{-1}$. Two subgroups $H_1$ and $H_2$ of $G$ are conjugate if there exists an element $y\in G$
such that $H_1=yH_2y^{-1}$.
For  any two elements $g,h\in G$ we denote $[g,h]$ the commutator
of $g$ and $h$, i.e.~$[g,h]=ghg^{-1}h^{-1}$. 
The commutator subgroup of $G$ is defined as $G'=\gen{[g,h]\st g,h\in G}$. 
The derived series of $G$ is defined recursively as $G^{(0)}=G$ and
$G^{(i+1)}=(G^{(i)})'$. The group $G$ is said to be solvable if there exists some integer $k$ such 
that $G^{(k)}=\triv$.
Given two groups $G_1$ and $G_2$, a map $\phi:G_1\to G_2$ is a homomorphism from $G_1$ to $G_2$ if, for any two elements 
$g$ and $g'$ in $G_1$, the relation $\phi(gg')=\phi(g)\phi(g')$ holds.  We say that $G_1$ and $G_2$ are isomorphic is there exists a 
one-one homomorphism from $G_1$ to $G_2$, and we write $G_1\cong G_2$.

Given a prime $p$, a $p$-group is a group of order $p^r$ for some integer $r$.
It is well-known that any $p$-group is solvable.
If $G$ is a group and $\abs{G}=p_1^{e_i}\ldots p_r^{e_r}$ for distinct prime numbers $p_i$ such that $p_1<\cdots<p_r$, then 
for each $i\bound{1}{r}$ the group $G$ has a subgroup  of order $p_i^{e_i}$ called a Sylow $p_i$-subgroup of $G$.
Moreover, if $G$ is additionally abelian, then each Sylow $p_i$-group is unique and
$G$ is the direct product of its Sylow subgroups.
Abelian $p$-groups have remarkably simple structures: any abelian 
$p$-group $P$ is isomorphic to a direct product of cyclic $p$-groups $\Int_{p^{e_1}}\times\cdots\times\Int_{p^{e_s}}$
for some positive integer $s$ and positive integers $e_1\le \ldots\le e_s$, and this decomposition is unique.
A total order $\preceq$ over the set of prime powers can be defined as follows: 
for any two prime powers $p^\alpha$ and $q^\beta$ where $\alpha$ and $\beta$ are positive integers, we write $p^{\alpha}\preceq q^{\beta}$ if and only if ($p<q$) or ($p=q$ and $\alpha\le \beta$).
We say that a list $(g_1,\ldots,g_t)$ of $t$ elements in $G$ is a basis of an abelian group $G$ if 
$G=\gen{g_1}\times\cdots\times\gen{g_t}$, the order of each $g_i$ is a prime power and $\abs{g_i}\preceq\abs{g_j}$ for any $1\le i\le j\le n$. 
It is easy to show that any (finite) abelian group has a basis and that, if 
$(g_1,\ldots,g_t)$ and $(g'_1,\ldots,g'_{t'})$ are two bases of $G$, then 
$t=t'$ and $\abs{g_i}=\abs{g'_i}$ for each $i\bound{1}{t}$.
For example, $(g_1,\ldots,g_t)$ is a basis of $G\cong \Int_2\times\Int_4\times\Int_3^2$ if and only if 
$t=4$, $\abs{g_1}=2$, $\abs{g_2}=4$, $\abs{g_3}=\abs{g_4}=3$ and $G=\gen{g_1}\times\gen{g_2}\times\gen{g_3}\times\gen{g_4}$.

Let $n$ be a positive integer. 
A Hall divisor of $n$ is a positive integer $m$ dividing $n$ such that 
$m$ is coprime with $n/m$. A subgroup $H$ of a finite group $G$ is called a Hall subgroup of $G$ if
$\abs{H}$ is a Hall divisor of $\abs{G}$. We will use in this paper the following well-known theorem.

\begin{theorem}[Hall's theorem]\label{theorem_hall}
Let $G$ be a finite solvable group and $r$ be a Hall divisor of $\abs{G}$. 
If $H_1$ and $H_2$ are two subgroups of $G$ with $\abs{H_1}=\abs{H_2}=r$, then $H_1$ and $H_2$ are conjugate.
\end{theorem}

We say that a finite group $G$ is an extension of a group $K$ by a group $L$ 
if there exists a normal abelian subgroup $N\cong K$ of $G$ such that $G/N\cong L$. 
We say that such an extension splits if there exists some subgroup $M$ of $G$ such that 
$G=NM$ and $N\cap M=\triv$. 
The Schur-Zassenhaus theorem states that any extension of $K$ by $L$ 
such that $gcd(\abs{K},\abs{L})=1$ splits. 
Split extensions can be constructed using the concept of semidirect products. 
Given two finite groups $K$ and $L$, and a homomorphism 
$\phi: L\to Aut(K)$, where $Aut(K)$ denotes the group of automorphisms of $K$, the 
semidirect product group $K\rtimes_\phi L$ is the group with ground set
$\{(k,\ell)\st k\in K,\ell\in L\}$ and group product 
$(k_1,\ell_1)\circ(k_2,\ell_2)=(k_1\phi(\ell_1)(k_2),\ell_1\ell_2)$. The direct product is the
special case where $\phi$ is the trivial homomorphism.
It is well known that a group is a split extension of $K$ by $L$ if and only if 
it is isomorphic to the semidirect product $K\rtimes_\phi L$ for some homomorphism $\phi:L\to Aut(K)$.
We usually drop the subscript of $\rtimes_\phi$ in the notation of the semidirect product 
when $\phi$ is implicit or not important. Thus an equivalent definition for the class $\Sp$ 
is the following: a group $G$ is in $\Sp$ if and only if there exist an abelian group $A$ and a cyclic group 
$\Int_{m}$ with $gcd(\abs{A},m)=1$ such that $G= A\rtimes \Int_m$.

In this paper we work in the black-box setting first introduced in \cite{Babai+FOCS84}.
A black-box group is a representation  of a group where elements are represented by 
strings (of the same length). 
An oracle that performs the group product is available: 
given two strings representing two elements $g$ and $g'$, 
the oracle outputs  the string representing $g\cdot g'$. Another oracle that, 
given a string representing an element $g$, computes a string representing 
the inverse $g^{-1}$ is available as well.
In this paper we assume the usual unique encoding hypothesis, i.e.~any element 
of the group is encoded by a unique string. 
We say that a group $G$ is input as a black-box if a set of strings representing 
generators $\{g_1,\ldots,g_s\}$ of $G$ with $s=O(\log\abs{G})$ is given as input,
and queries to the multiplication and inversion oracles can be done at cost 1.
The hypothesis on $s$ is natural since every group $G$ has a generating set of size $O(\log \abs{G})$, and enables us 
to make the exposition of our results easier. 
The complexity of our algorithm (the bound $(\sqrt{n}+\gamma)^{1+o(1)}$ in Theorem \ref{theorem_main}) is unchanged if $s=\abs{G}^{o(1)}$ and is 
still polynomial for any larger value of $s$. Also notice that a set of generators of any size can 
be converted efficiently into a set of generators of size $O(\log\abs{G})$ if randomization 
is allowed \cite{BabaiSTOC91}.

\section{Computing a Standard Decomposition}\label{section_standard}
For a given group $G$ in the class $\Sp$ in general many different decompositions as a semidirect product of an abelian group by a 
cyclic group exist. For example, the abelian group $\Int_6=\gen{x_1,x_2 \st x_1^2=x_2^3=[x_1,x_2]=e}$ 
can be written as $\gen{x_1}\times\gen{x_2}$, $\gen{x_2}\times\gen{x_1}$ or $\gen{x_1,x_2}\times\triv$.
That is why we introduce the notion of a standard decomposition. 
Let us first start with a simple definition. 
\begin{definition}
Let $G$ be a finite group. 
For any positive integer $m$ denote by $\Dd^m_G$ the set (possibly empty) of pairs $(A,B)$ such that the following
three conditions hold:
(i)
$A$ is a normal abelian subgroup of $G$ of order coprime with $m$; and
 (ii)
 $B$ is a cyclic subgroup of $G$ of order $m$; and
 (iii)
 $G=AB$.
 \end{definition}
 Notice that if for some $m$ the set $\Dd^m_G$ is not empty, then $G$ is in the class $\Sp$.
 Conversely, if $G$ is in $\Sp$, then there exists at least one integer $m$ 
 such that $\Dd_G^m$ is not empty. 
Also notice that $\gamma(G)$ is the smallest positive integer such that $\Dd^{\gamma(G)}_G\neq\varnothing$.
We now define the concept of a standard decomposition.
 \begin{definition}
Let $G$ be a group in the class $\Sp$.
A standard decomposition of $G$ is an element of $\Dd^{\gamma(G)}_G$.
 \end{definition}
\noindent The following simple lemma will be very  useful.
 \begin{lemma}\label{lemma_Hall}
Let $G$ be a group in $\Sp$ and $m$ be any positive integer. 
If $(A_1,B_1)$ and $(A_2,B_2)$ are two elements of $\Dd^m_G$, then $A_1=A_2$.
\end{lemma}
\proof
Let us write $B_1=\gen{y_1}$.
Any element $g$ of $A_2$ can be written as $g=hy_1^c$ with $h\in A_1$ and some integer $c$. If $c\not\equiv 0\bmod m$,
then $gcd(m,\abs{g})\neq 1$, which is excluded since $\abs{A_2}$ and $m$ are coprime. 
Then $A_2\subseteq A_1$. By symmetry $A_1\subseteq A_2$ and $A_1=A_2$.
\qed

Before explaining how to compute a standard definition for a group in $\Sp$, let us mention that it 
is well known that the order of an element $g$ of any finite group $G$ can be computed deterministically in time
$\tilde O(\abs{G}^{1/2})$ using Shanks' baby-step/giant-step method \cite{Shanks69} or its variants 
\cite{Shoup05}. 
Here, for two functions $f$ and $g$ with appropriate domains and ranges, the notation $g(n)=\tilde O(f(n))$ 
means that there exists a positive constant $c$ such that $g(n)=O(f(n)(\log(f(n)))^c)$. 
In the following proposition we show that the decomposition of an element in an abelian group can be 
found efficiently by a very similar approach
(we will need this in Section \ref{section_algorithm}).
\begin{proposition}\label{proposition_abmem}
Let $A$ be an abelian group and $(g_1,\ldots,g_s)$ be a basis of $A$. There exists a deterministic 
algorithm with time complexity $\tilde O(\abs{A}^{1/2})$ that, given any element $g\in A$, outputs integers
$a_1,\ldots,a_s$ such that $g=g_1^{a_1}\cdots g_s^{a_s}$.
\end{proposition}
\proof
Denote $r_i=\sqrt{\abs{g_i}}$ for each $i\bound{1}{s}$ and, for simplicity, suppose that $r_i$ is an integer. The case where $r_i$
is not an integer is similar. 
The algorithm first computes the set $S=\{g_1^{c_1}\cdots g_s^{c_s} \st c_i\bound{0}{r_i-1}\}$.
Then the algorithm tries all the elements $(b_1,\ldots,b_s)$ with $b_i\bound{0}{r_i-1}$ until finding
an element $(\bar b_1,\ldots,\bar b_s)$ such that $gg_1^{-\bar b_1 r_1}\cdots g_s^{-\bar b_s r_s}\in S$. 
Denote $gg_1^{-\bar b_1 r_1}\cdots g_s^{-\bar b_sr_s}=g_1^{c_1}\cdots g_s^{c_s}$, where each $c_i$ is
an element of $\{1,\ldots,r_i-1\}$. A clever way for finding the $c_i$'s is to use an appropriate data 
structure for storing $S$. Then the algorithm outputs $(r_1\bar b_1+c_1,\ldots,r_s \bar b_s+c_s)$.
The correctness of this algorithm follows immediately from the fact that, if $g=g_1^{a_1}\cdots g_s^{a_s}$, then
each $a_i$ can be written as $a_i=\bar b_i r_i+c_i$ for some $\bar b_i$ and $c_i$ in $\{0,\ldots,r_i-1\}$. Its complexity
is $\tilde O(\abs{A}^{1/2})$.\qed

We now show how to compute a standard decomposition of any group in the class $\Sp$ in time polynomial in the order of the group.
The key part of the algorithm is the following procedure $\proc{Find-Decomposition}$ that,
given a group $G$ in $\Sp$ and an integer $m$, computes an element of  $\Dd^m_G$ if this set
is not empty. 
The description is given in metacode, followed by more details. \vspace{0mm}

\begin{codebox}
\Procname{Procedure $\proc{Find-Decomposition}(G,m)$} 
\zi \const{input:} a set of generators $\{g_1,\ldots,g_s\}$ of a group $G$ in $\Sp$ with $s=O(\log\abs{G})$ 
\zi \hspace{12mm} a positive integer $m$ dividing $\abs{G}$
\zi \const{output:} an error message or a pair $(M,z)$ where $z\in G$ and $M$ is a subset of $G$
\li compute a set of generators $\{x_1,\ldots,x_t\}$ of $G'$ with $t=O(\log\abs{G})$;
\li factorize $m$ and write $m=p_1^{e_1}\cdots p_r^{e_r}$;
\li search indexes $k_1,\ldots,k_r\in\{1,\ldots,s\}$ such that $p^{e_\ell}_\ell$ divides $\abs{g_{k_\ell}}$ for each $1\le \ell\le r$;
\li \If no such $r$-uple $(k_1,\ldots,k_r)$ exists 
\li \Then return  \const{error}; 
\li \Else
\li   $g\gets \Pi_{\ell=1}^r g_{k_{\ell}}^{\abs{g_{k_{\ell}}}/p_\ell^{e_\ell}}$;
\li \If $m$ does not divide $\abs{g}$ 
\li \Then return \const{error};  
\li \Else
\li   $z\gets g^{\abs{g}/m}$;
\li   \For $j =1$ \To $s$ \kw{do} $h_j\gets g_j^m$;     
\li \If $\gen{x_1,\ldots,x_t,h_1,\ldots,h_s}$ is abelian 
\zi \hspace{20mm}and $gcd(\abs{x_i},m)=1$ for each $i\bound{1}{t}$  
\zi \hspace{20mm}and $gcd(\abs{h_\ell},m)=1$ for each $\ell\bound{1}{s}$ 
\li \Then  return $(\{x_1,\ldots,x_t,h_1,\ldots,h_s\},z)$; 
\li \Else return \const{error};  
\End
\li \hspace{-8mm}\bf{endelse}
\End
\li \hspace{-8mm}\bf{endelse}
\end{codebox}\vspace{0mm}

At Step 1 a set of generators $\{z_1,\ldots,z_{t'}\}$ of $G'$ with $t'=O(s^3)$ can be computed using 
$O(s^3)$ group operations by noticing that $G'=\gen{g_k[g_i,g_j]g_k^{-1}\st i,j,k\bound{1}{s}}$ 
(we refer to \cite{Holt+05} for a proof of this simple fact). Since $G'$ is abelian for any
group $G$ in the class $\Sp$, a generating set $\{x_1,\ldots,x_t\}$ of $G'$ with $t=O(\log\abs{G})$ 
can then be obtained
in time $\tilde O(\abs{G}^{1/2})$
using the deterministic algorithm by Buchmann and Schmidt \cite{Buchmann+05} 
that computes a basis of any abelian group
$K$ in time $\tilde O(\abs{K}^{1/2})$. 
At Step 2 the naive technique for factoring $m$ (trying all the integers up to $\sqrt{m}$) is sufficient. 
This takes $\tilde O(\abs{G}^{1/2})$ time. 
At Steps 3, 7 and 13 we use Shanks' method \cite{Shanks69} to compute orders of elements of
$G$ in time $\tilde O(\abs{G}^{1/2})$.
At step 13, commutativity is tested by checking that every two generators commute: this can be done in 
$O(s^2+t^2)$ group operations.  
Proposition \ref{proposition_correctness1} below summarizes the time complexity of the procedure and prove its correctness.
We state first one simple lemma.
\begin{lemma}\label{lemma_power}
Let $G$ be a group in $\Sp$ and $(A,B)$ be a standard decomposition of $G$. Denote $\abs{B}=m$.
Let $\{g_1,\ldots,g_s\}$ be a set of generators of $G$. Then $A=\gen{G',g_1^m,\ldots,g_s^m}$,
where $G'$ is the derived subgroup of $G$.
\end{lemma}
\proof
Let $B=\gen{y}$ and,  for each $i\bound{1}{s}$,  write $g_i$ as $z_iy^{k_i}$ for some $z_i\in A$ and $k_i\bound{1}{m}$.
Then $A=\gen{G',z_1,\ldots,z_s}$. Notice that $G'$ has to be included since in general 
$A\neq\gen{z_1,\ldots,z_s}$, 
e.g.~$G=\gen{x_1,x_2,y\st x_1^3=x_2^3=y^2=e,yx_1=x_2y,yx_2=x_1y}$ with the generating set 
$g_1=x_1y$ and $g_2=y$. A simple computation shows that $g_i^m=u_iz_i^my^{mk_i}=u_iz_i^m$
for some element $u_i\in G'$. Since $m$ is coprime with the order of $z_i$, we conclude that 
$A=\gen{G',g_1^m,\ldots,g_s^m}$.
\qed

\begin{proposition}\label{proposition_correctness1}
The time complexity of the procedure $\proc{Find-Decomposition}(G,m)$ is 
$\tilde O(\abs{G}^{1/2})$.
If $\Dd^m_G\neq\varnothing$, then $\proc{Find-Decomposition}(G,m)$ outputs a pair $(M,z)$ such that $(\gen{M},\gen{z}))\in\Dd^m_G$.
Conversely, if $\proc{Find-Decomposition}(G,m)$ does not output an error message, then its output $(M,z)$ is such 
that $\gen{M,z}\in \Sp$ and $(\gen{M},\gen{z})\in\Dd^m_{\gen{M,z}}$. 
\end{proposition}
\proof
It is clear that the procedure always terminates since no loop is used. 
The time complexity follows from the analysis of Steps 1, 2, 3, 7 and 13 already done, and from the fact that
$s=O(\log\abs{G})$.

Suppose that $\Dd^m_G\neq\varnothing$ and take a decomposition $(A,\gen{y})\in\Dd^m_G$. 
Write $m=p_1^{e_1}\cdots p_r^{e_r}$ for primes $p_1<\cdots<p_r$ and denote $q_\ell=p_\ell^{e_\ell}$
for each $\ell\bound{1}{r}$.
Notice that for any generating set $\{g_1,\ldots,g_s\}$ of $G$,  and for each $\ell\bound{1}{r}$, 
there should be some index $k_\ell$ for which $g_{k_\ell}$ is of the form 
 ${u_\ell}y^{c_\ell}$, where $u_\ell\in A$ and $c_\ell$ is such that $q_\ell$ divides the order of $y^{c_\ell}$,
 i.e.~$q_\ell$ divides $m/gcd(m,c_\ell)$. 
Also notice that in this case $q_\ell$ divides the order of $g_{k_\ell}$ as well.  
Then the element $\bar{g}_{k_\ell}=g_{k_{\ell}}^{\abs{g_{k_{\ell}}}/q_\ell}$
has order $q_\ell$ and, more precisely, is of the form $v_\ell y^{d_\ell}$
for some $v_\ell\in A$ and some $d_\ell=\gamma_\ell m/q_\ell$ with $\gamma_\ell$ coprime with $m$.
Then the element $g=\Pi_{\ell=1}^r \bar{g}_{k_{\ell}}$ is of the form 
$wy^{d}$ where $w\in A$ and $d=d_1+\cdots+d_r$ is coprime with $m$. 
Thus $m$ divides $\abs{g}$ and $z=g^{\abs{g}/m}$ is an element of order $m$ of the form $w'y^e$ 
with $e$ coprime with $m$. 
From Lemma \ref{lemma_power} we know that $\gen{x_1,\ldots,x_t,h_1,\ldots,h_s}=A$
and conclude that $(\gen{x_1,\ldots,x_t,h_1,\ldots,h_s},\gen{z})\in \Dd^m_G$. 

We now prove the last part of the proposition.
Suppose that the algorithm does not err and denote $(M,z)$ its output. 
Then $z$ has order $m$ and $\gen{M}$ is an abelian subgroup of $G$ of order coprime with $m$,
since the tests at steps 8 and 13 succeeded.
Moreover $\gen{M}$ is normal in $G$ since $G'\le \gen{M}$. We conclude that 
$\gen{M,z}\in \Sp$ and
$(\gen{M},\gen{z})\in\Dd^m_{\gen{M,z}}$.
\qed

We now present an algorithm computing a standard decomposition of any group in $\Sp$.

\begin{theorem}\label{theorem_basis}
There exists a deterministic algorithm that, on an input $G$ in the class $\Sp$ given as a black box,
outputs  an element $z\in G$ and a set $M$ of elements in $G$ 
such that  $(\gen{M},\gen{z})$ is a standard decomposition of $G$.
The time complexity of this algorithm is $O(\abs{G}^{1/2+o(1)})$.
\end{theorem}
\proof
The algorithm is as follows. Let $G$ be a group in the class $\Sp$, input as a black box with generating set
$\{g_1,\ldots,g_s\}$ where $s=O(\log\abs{G})$.

We first compute $\abs{g_i}$ for each $i\bound{1}{s}$ using Shanks' algorithm. 
Let $\bar{m}$ be the least common multiple of the $s$ integers $\abs{g_1},\ldots,\abs{g_s}$.
We compute the set $S$ of divisors of $\bar{m}$, and denote $m_1<m_2<\cdots<m_r$ the elements of $S$ in increasing order.
 
 For $i$ from 1 to $r$ we run the procedure $\proc{Find-Decomposition}(G,m_i)$ on the set 
 $\{g_1,\ldots,g_s\}$ and $m_i$, and obtain an error message or an output $(\gen{M_i},z_i)$.
 Let $n$ be the maximum value of the quantity $m_i\abs{\gen{M_i}}$ over all the $i$'s such that the output is not an error message
 (we will show that for at least one value of $i$ the output is not an error message so $n$ is well defined).
Notice that computing $\abs{M_i}$ can be done using the deterministic algorithm by Buchmann and Schmidt \cite{Buchmann+05}
that computes the order of any abelian group $K$ in time $\tilde O(\abs{K}^{1/2})$.
 Finally the algorithm takes the smallest integer $i_0\bound{1}{r}$ such that $m_{i_0}\abs{M_{i_0}}=n$, and then outputs $z_{i_0}$ and $M_{i_0}$.  
 
We now analyze this algorithm. First of all notice that for any $m$ such that $\Dd^{m}_G$ is not empty, 
this integer $m$ is in $S$ since $m$ divides $\bar{m}$. 
By Proposition $\ref{proposition_correctness1}$, if  $\Dd^{m_i}_G$ is not empty then 
the procedure 
$\proc{Find-Decomposition}(G,m_i)$ outputs an element $(\gen{M_i},\gen{z_i})\in\Dd^{m_i}_G$ 
and then $m_i\abs{\gen{M_i}}=\abs{G}$. 
Conversely, and again by Proposition $\ref{proposition_correctness1}$, 
if the procedure 
$\proc{Find-Decomposition}(G, m_i)$ outputs $(M_i,z_i)$, then $m_i\abs{\gen{M}}=\abs{\gen{z_i,M_i}}\le \abs{G}$.
Thus $n$ is well defined and is equal to the order of $G$. 
Finally, trying all the elements of $S$ gives clearly 
the minimal $m$ such that $\Dd^{m}_G$ is not empty. Then $(\gen{M_{i_0}},z_{i_0})$ is a standard 
decomposition of $G$.
The time complexity of the algorithm is shown to be $\abs{G}^{1/2+o(1)}$ using Proposition \ref{proposition_correctness1} and the following two facts.
First, computing the set $S$ can be done in $\tilde O(\abs{G}^{1/2})$ time.
Second, the number of divisors of any integer $k$ has for upper bound  $O(k^\varepsilon$) for any positive constant
 $\varepsilon$ (see for example \cite{HardyWright79}). Since $\bar{m}\le \abs{G}$ we conclude that $r=\abs{G}^{o(1)}$.
\qed

\noindent{\bf Remark.}
The space complexity of the algorithm of Theorem \ref{theorem_basis}
 is $\tilde \Theta(\sqrt{\abs{G}})$ since the baby-step/giant-step method 
requires this amount of space. An algorithm working in space polynomial in 
$\log\abs{G}$ can also be 
constructed but in this case the time complexity gets worse 
(but is still polynomial in $\abs{G}$).

\section{Testing Conjugacy}\label{section_conjugacy}
In this section we study the automorphism group of any abelian group and describe how to decide 
whether two automorphisms are conjugate.

Let $A$ be a finite abelian group. Then $A$ is the direct product of all its Sylow subgroups.
Since $Aut(A)$ is the direct product of the automorphism groups of the Sylow subgroups, 
we can assume without loss of generality that $A$ is an abelian $p$-group for some prime $p$.
In this section we suppose that  $A$ is isomorphic to the group $\Int_{p^{e_1}}\times\cdots\times\Int_{p^{e_s}}$,
for some positive integers $s$ and $e_1\le e_2\le \ldots\le e_s$. 
\subsection{Automorphisms of an abelian group}\label{subsection_auto}
We first introduce a matricial characterization of the automorphism group of $A$, 
following the work of Ranum \cite{Ranum07}.

Let $(g_1,\ldots,g_s)$ be 
a basis of $A$, i.e.~$s$ elements of $A$ such that the order of each $g_i$ is $p^{e_i}$ and such that 
$A=\gen{g_1}\times\cdots\times\gen{g_s}$. 
Let $\psi$ be an endomorphism of $A$ and, for each $j\in\{1,\ldots,s\}$, 
denote $\psi(g_j)=g_1^{u_{1j}}\dots g_s^{u_{sj}}$ where each $u_{ij}$ is in the set 
$\{0,\ldots,p^{e_i}-1\}$. 
The values $u_{ij}$, which can be seen as an integer matrix $(u_{ij})$ of size $s\times s$, fully define the endomorphism $\psi$.
However the converse is not true: an arbitrary integer matrix $(u_{ij})$ of size $s\times s$ with each value $u_{ij}$ in  $\{0,\ldots,p^{e_i}-1\}$
does not necessarily define an endomorphism of $A$, because $\psi$ should be a homomorphism, and not only a linear map. 
It is easy to give necessary and sufficient conditions for these values $u_{ij}$ to define an endomorphism of $A$: 
$p^{e_i-e_{min(i,j)}}$ should divide $u_{ij}$ for any $i,j\bound{1}{s}$.

\begin{definition}\label{theo}
Define $M(A)$ as the following set of integer matrices.
$$M(A)=\left\{
(u_{ij})\in \Int^{s\times s} \st 0\le u_{ij}< p^{e_i} \textrm{ and }  p^{e_i-e_{min(i,j)}} \textrm{ divides } u_{ij} \textrm{ for all } i,j\bound{1}{s}
\right\}
$$
Given $U$ and $U'$ in $M(A)$ define the  multiplication $\ast$ as follows: 
$U\ast U'$ is the integer matrix $W$ of size $s\times s$ such that 
$w_{ij}=(\sum_{k=1}^s u_{ik}u'_{kj} \bmod p^{e_i})$ for $i,j\bound{1}{s}$, i.e.~after computing
the usual matrix multiplication $UU'$, each entry is reduced modulo $p^{e_i}$, where $i$ is the
row of the entry. 
Let $R(A)$ be the set
$
R(A)=\left\{U\in M(A) \st det(U)\not\equiv 0 \bmod p
\right\}.
$
\end{definition}
Ranum has shown that the set $R(A)$ corresponds to the set of automorphisms of $A$ \cite{Ranum07}.   

\begin{theorem}{(\cite{Ranum07})}
The set $R(A)$ with the product operation $\ast$ is a group isomorphic to the group
of automorphisms of $A$.
\end{theorem}

Let us consider a few important examples to illustrate the definitions introduced.@\vspace{3mm}

\noindent{\bf Example 4.1.}
If $A=\Int_p^s$ for some integer $s$, then $M(A)$ is the set of matrices of size $s\times s$ over the finite field
$\Int_p$ and $R(A)$ is the general linear group $GL_s(p)$ of invertible matrices of size $s\times s$ over $\Int_p$.\vspace{3mm}

\noindent{\bf Example 4.2.} 
Let $A$ be the group $\Int_{p}\times\Int_{p^2}\times\Int_{p^2}\times\Int_{p^5}$, then
\begin{equation}\label{eq_example}
M(A)=
\left\{\left(\begin{array}{llll}
\lambda_{11}&\lambda_{12}&\lambda_{13}&\lambda_{14}\\
p\lambda_{21}&\lambda_{22}&\lambda_{23}&\lambda_{24}\\
p\lambda_{31}&\lambda_{32}&\lambda_{33}&\lambda_{34}\\
p^4\lambda_{41}&p^3\lambda_{42}&p^3\lambda_{43}&\lambda_{44}
\end{array}\right)
\st 0\le \lambda_{ij}<p^{e_{min(i,j)}}
\right\}.
\end{equation}
\subsection{Structure of the automorphism group}
We analyze now in more details the structure of the group $R(A)$.
Several new definitions are introduced and we refer to the end of this subsection for an example.

We write $A\cong H_1\times\cdots\times H_t$ with $H_i=\Int_{p^{f_i}}^{k_i}$ 
where $f_1<f_2<\cdots<f_t$ are positive strictly increasing integers and $k_1,\ldots,k_t$ are positive integers. 
Notice that $t$ and these integers are uniquely determined. In particular $f_i$ is the $i$-th smallest 
element in the series $(e_1,\ldots,e_s)$ and $k_i$ is the number of times $f_i$ appears in the series. 
Also notice that $k_1+\cdots+k_t=s$.
 
Let $U=(u_{ij})$ be an element of $M(A)$. We define $t$ blocks $D_1(U),\ldots,D_t(U)$ of $U$ as follows: 
$D_i(U)$ is the matrix of size $k_i\times k_i$ obtained by selecting the rows and columns with indexes from 
$(k_1+\cdots+k_{i-1}+1)$ to $(k_1+\cdots +k_{i-1}+k_i)$. Notice that $D_i(U)$ is a matrix in $M(H_i)$ which 
lies on the diagonal of $U$. For any matrix $U$ in $M(A)$ and any $i\bound{1}{t}$, 
denote $[U]_i$ the matrix  obtained by reducing the entries of $D_i(M)$ modulo $p$. Each $[U]_i$ can then be seen
as an element of $GL_{k_i}(p)$.
For each $i\bound{1}{t}$ we also define the following subset of $M(H_i)$.
$$K_i(A)=\left\{(u_{ij})\in M(H_i)\st p \textrm{ divides } (u_{ij}-\delta_{ij}) \textrm { for all } i,j\bound{1}{k_i}\right\}.
$$
In the definition of $K_i(A)$, $\delta_{ij}$ is the Kronecker symbol (equal to 1 if $i=j$ and equal to 0 otherwise).
In other words, each diagonal entry of a matrix in $K_i(A)$ is of the form $1+p\lambda_{ii}$ and each non-diagonal entry
is of the form $p\lambda_{ij}$. Finally we introduce the following definition.
\begin{definition}
Consider the subset $N(A)$ of $M(A)$ defined as follows.
$$N(A)=\left\{U\in M(A) \st D_i(U)\in K_i(A)\textrm{ for each } i\bound{1}{t}\right\}$$
Also consider the subgroup $V(A)$ of the group $GL_s(p)$ defined as
$$
V(A)=\left\{
A\in GL_s(p) \st V=diag(V_1,\ldots,V_t)\ \textrm{ with } V_i\in GL_{k_i}(p)\textrm{ for each } i\bound{1}{t}
\right\}.
$$
Let $\Psi$ be the map from $R(A)$ to $V(A)$ such that  $\Psi(U)=diag([U]_1,\ldots,[U]_t)$ for any $U\in R(A)$, i.e.~the
diagonal blocks of $A$ are reduced modulo $p$ and the others entries are mapped to zero. 
\end{definition}

We now prove the following result.
\begin{proposition}\label{prop_surj}
$\Psi$ is a surjective group homomorphism from $R(A)$ to V(A). Its kernel is $N(A)$.
\end{proposition}
\begin{proof}
$\Psi$ is clearly surjective and $\Psi^{-1}(I)=N(A)$ where $I$ denotes the identity of $V(A)$.
Take two arbitrary matrices $U$ and $U'$ in $R(A)$.
To prove that $\Psi$ is an homomorphism we have only to prove that $[U\ast U']_i=[U]_i[U']_i$ for
each $i\bound{1}{s}$. This is easy to show by noticing that all the entries on the left and below the 
diagonal blocks of $U$ and $U'$ are divided by $p$.
\end{proof}
Proposition \ref{prop_surj} shows 
that $N(A)$ is a normal subgroup of $R(A)$ and $R(A)/N(A)\cong V(A)$.\vspace{4mm}

\noindent{\bf Example 4.3.} 
Let $A$ be again the group $\Int_{p}\times\Int_{p^2}\times\Int_{p^2}\times\Int_{p^5}$.
Then $t=3$, $f_1=1$, $f_2=2$, $f_3=5$, $k_1=k_3=1$ and $k_2=2$, i.e.~$H_1=\Int_p$, 
$H_2=\Int_{p^2}^2$ and $H_3=\Int_{p^5}$. Then, using the notation for a general 
element $U$ in $M(A)$ used in Equation (\ref{eq_example}) we obtain
$D_1(U)=(\lambda_{11})$,
$
D_2(U)=
\left(\begin{array}{ll}
\lambda_{22}&\lambda_{23}\\
\lambda_{32}&\lambda_{33}
\end{array}\right),
$
and
$
D_3(U)=(\lambda_{44})$.
The sets $K_i(A)$ are as follows:
$K_1(A)=\{(1)\}$,
$K_2(A)=
\left\{
\left(\begin{array}{cc}
1+p\alpha_{11}&p\alpha_{12}\\
p\alpha_{21}&1+p\alpha_{22}
\end{array}\right)
\st 0 \le \alpha_{ij}<p 
\right\}$, 
and $K_3(A)=\{(1+p\alpha)\st 0\le \alpha<p^4\}$.
We conclude that

\begin{equation}\label{eq_kernel}
N(A)=
\left\{\left(\begin{array}{c|c|c}
C_1&
\begin{array}{cc}\lambda_{12}&\lambda_{13}\end{array}&
\lambda_{14}\\
\hline
\begin{array}{c}p\lambda_{21}\\p\lambda_{31}\end{array}&
{\LARGE C_2}&
\begin{array}{c}\lambda_{24}\\\lambda_{34}\end{array}\\
\hline
p^4\lambda_{41}&
\begin{array}{cc}p^3\lambda_{42}&p^3\lambda_{43}\end{array}&
C_3
\end{array}\right)
\st 0\le \lambda_{ij}<p^{e_{min(i,j)}}, C_k\in K_k(A)
\right\}.
\end{equation}
Then $V(A)$ is the set of matrices of the form $diag(V_1,V_2,V_3)$
where $V_1,V_3\in GL_1(p)$ and $V_2\in GL_2(p)$.
Finally we give an example of the action of $\Psi$ (suppose here that $p\neq 2$):
\[
\Psi: 
\left(\begin{array}{c|cc|c}
2&1&3&p\\
\hline
3p&1&p+2&p+1\\
p&p+1&p&2\\
\hline
3p^4&p^3&2p^3&p^2+1
\end{array}\right)
\mapsto
\left(\begin{array}{c|cc|c}
2&0&0&0\\
\hline
0&1&2&0\\
0&1&0&0\\
\hline
0&0&0&1
\end{array}\right).
\]

\subsection{Testing conjugacy in $R(A)$}

In this subsection we consider the following computational problem and present an efficient algorithm solving it.\vspace{3mm}

$\proc{Conjugacy}$\\\vspace{-5mm}

\const{input:}
an abelian $p$-group $A$ and two matrices $U_1$ and $U_2$ in $R(A)$ such that \vspace{-2mm}
\begin{equation}\label{condition}\vspace{-2mm}
the \: orders \: of\:U_1\: and\:U_2\: are\: coprime\: with\:  p 
\end{equation}

\const{output:} an element $U\in R(A)$ such that $U\ast U_1= U_2\ast U$ if such an element exists\vspace{4mm}

The problem $\proc{Conjugacy}$ asks to check whether two matrices $U_1$ and $U_2$ in $R(A)$ satisfying 
condition (\ref{condition}) are conjugate in $R(A)$.
Trying all the possibilities for $U$ requires $\abs{R(A)}$ trials. Since for example in the case 
$A=\Integers_{p^k}^s$ with $p$ and $k$ constant the bound $\abs{R(A)}=\Theta(\abs{A}^{\log\abs{A}})$ holds, such a naive approach is not efficient. 
However, notice that in the case $A=\Int_{p}^s$ the group $A$ has more than the structure of an abelian group: $A$ is a vector space
over the field $\Int_p$ and then $R(A)=GL_s(p)$. A mathematical criterion for the conjugacy of matrices in 
$GL_s(p)$ (even without the condition (\ref{condition}) on their orders) is known: two matrices are conjugate if and only if their
canonical rational forms are equal. Since the canonical rational form of a matrix can be computed efficiently \cite{SteelJSymb97},
this gives an algorithm solving the problem $\proc{Conjugacy}$ in time polynomial in $\log\abs{A}$. 
However, when $A$ has no vector space structure, there is no known simple mathematical criterion for the conjugacy 
of matrices and, to our knowledge, no algorithm faster than  the above naive approach is known, even for the case where $A=\Int_{p^2}^s$.
We now show that with the additional condition (\ref{condition}) on the order of $U_1$ and $U_2$ there exists 
an algorithm solving the problem $\proc{Conjugacy}$ in time polynomial in $\log{\abs{A}}$ for any abelian $p$-group $A$.

Our algorithm is based on the following proposition, which is a generalization of an argument by Pomfret \cite{Pomfret73}.
\begin{proposition}\label{proposition_conj}
Let $A$ be an abelian $p$-group and $U_1,U_2$ be two matrices in $R(A)$ of order coprime with $p$. 
Then $U_1$ and $U_2$ are conjugate in $R(A)$ if and only if $\Psi(U_1)$ and $\Psi(U_2)$ are conjugate in $V(A)$.
Moreover if $U_1$ and $U_2$ are conjugate in $R(A)$ then for any $X\in R(A)$ such that 
$\Psi(U_1)=\Psi(X)^{-1}\Psi(U_2)\Psi(X)$ there exists a matrix $Y\in N(A)$ such that $X\ast Y\ast U_1= U_2\ast X\ast Y$.
\end{proposition}
\begin{proof}
For brevity we omit the symbol $\ast$ when denoting multiplications in $R(A)$.
Since $\Psi$ is an homomorphism, if $U_1$ and $U_2$ are conjugate in $R(A)$ then $\Psi(U_1)$ and $\Psi(U_2)$ are conjugate in $V(A)$.
Now suppose that $\Psi(U_1)$ and $\Psi(U_2)$ are conjugate in $V(A)$. Since the image of $\Psi$ is $V(A)$,
there exists some $X\in R(A)$ such that  $\Psi(U_1)=\Psi(X)^{-1}\Psi(U_2)\Psi(X)$ and thus $U_1=X^{-1}U_2XM$ for some $M\in N(A)$.
Then $\gen{U_1}N(A)=\gen{X^{-1}U_2X}N(A)$ (since $N(A)$ is a normal subgroup of $R(A)$) and the two subgroups $\gen{U_1}$ and $\gen{X^{-1}U_2X}$ are Hall subgroups of the group 
$\gen{U_1}N(A)$.
Moreover since $\gen{U_1}N(A)$ is a cyclic extension of the $p$-group $N(A)$, this is a solvable group.
Then, from Theorem \ref{theorem_hall}, this implies that the two subgroups  $\gen{U_1}$ and $\gen{X^{-1}U_2X}$ are conjugate in $\gen{U_1}N(A)$ and thus 
there exists an element $Y\in \gen{U_1}N(A)$ and some $r>0$ such that $Y^{-1}X^{-1}U_2XY=U_1^r$. 
Without loss of generality $Y$ can be taken in $N(A)$.
Thus $\Psi(U_1)=\Psi(X)^{-1}\Psi(U_2)\Psi(X)=\Psi(U_1)^r$. Since the order of the kernel of $\Psi$ is coprime with the order of $U_1$, 
the matrices $U_1$ and $\Psi(U_1)$ have the same order, and thus $U_1=U_1^r$. We conclude that $Y^{-1}X^{-1}U_2XY=U_1$. The matrices
$U_1$ and $U_2$ are thus conjugate in $R(A)$.
The second part of the theorem follows from the observation that $X$ can be chosen in an arbitrary way.
\end{proof}


We now present our algorithm.
\begin{theorem}\label{theorem_conj}
There exists a deterministic  algorithm that solves the problem $\proc{Conjugacy}$
in time polynomial in $\log\abs{A}$.
\end{theorem}
\begin{proof}
The algorithm is as follows.

Given $U_1$ and $U_2$ in $R(A)$ satisfying Condition (\ref{condition}), we first 
compute the two matrices $V_1=\Psi(U_1)$ and 
$V_2=\Psi(U_2)$ in $V(A)$. 
Then we check the conjugacy of $V_1$ and $V_2$ in $V(A)$ 
using the following approach. $V_1$ and $V_2$ are conjugate in $V(A)$ if and only if the blocks 
$D_i(V_1)$ and $D_i(V_2)$ are conjugate in $GL_{k_i}(p)$ for each $i\bound{1}{t}$, that is, if 
$D_i(V_1)$ and $D_i(V_2)$  have the same rational normal form.
The rational normal form of matrices of size $n\times n$ (and transformation matrices) over any finite field 
can be computed using $O(n^4)$ field operations (see for example \cite{SteelJSymb97}).
Thus we can decide in time polynomial in $\log\abs{A}$ whether 
$D_i(V_1)$ and $D_i(V_2)$ are conjugate for all $i\bound{1}{t}$.
If this is not the case then we conclude that $U_1$ and $U_2$ are not conjugate in $R(A)$
from Proposition \ref{proposition_conj}. Otherwise $U_1$ and $U_2$ are conjugate in $R(A)$
and the remaining of the proof shows how to compute a matrix $U\in R(A)$ such that $U\ast U_1=U_2\ast U$. 

We compute transformation matrices $T_i\in GL_{k_i}(p)$, for $i\bound{1}{t}$, such that $T_iD_i(V_1)=D_i(V_2)T_i$
using, for example, again the  algorithm \cite{SteelJSymb97}.
Then we take any matrix $X$ in $R(A)$ such that $\Psi(X)=diag(T_1,\ldots,T_t)$, e.g.~the
matrix $X$ in $R(A)$ with diagonal blocks equal to $T_1,\ldots,T_t$ and 
zero everywhere else. 
We finally determine a solution $Y$ in $N(A)$ of the matrix equation $X\ast Y\ast U_1=U_2\ast X\ast Y$. Such 
solution exists by Proposition \ref{proposition_conj}.
To do this, we write the general form of an element $Y$ of $N(A)$ using $s^2$ variables 
$y_{ij}$:  the entry corresponding to the $i$-th row and the $j$-th column of $Y$, for $i,j\bound{1}{s}$,
is of the form $(1+py_{ij})$ if $i=j$ and is of the form $p^{d_{ij}}y_{ij}$ for some appropriate nonnegative 
integer $d_{ij}$ otherwise (see Equation (\ref{eq_kernel}) for an example).
Then the equation $X\ast Y\ast U_1=U_2\ast X\ast Y$ can be rewritten as the following system of $s^2$ linear modular equations of 
$s^2$ variables $y_{ij}$: 
$$\sum_{i,j=1}^s\alpha_{ij}^{(k,\ell)}y_{ij}\equiv \beta^{(k,\ell)} \bmod p^{e_k} \textrm{ for } 1\le k,\ell\le s,$$
where $\alpha^{(k,\ell)}_{ij}$ and $\beta^{(k,\ell)}$ are known.
Now we add on each modular equation a new variable $z_{k\ell}$ with coefficient $p^{e_k}$. 
This transforms the above system into the following system of $s^2$ linear Diophantine solutions of 
$2s^2$ variables:
$$\sum_{i,j=1}^s\alpha_{ij}^{(k,\ell)}y_{ij}+p^{e_k} z_{k\ell}= \beta^{(k,\ell)} \textrm{ for } 1\le k,\ell\le s .$$
It is known that any system of linear Diophantine equations with $n_1$ equations and $n_2$ variables can be solved in time polynomial 
in $n_1$, $n_2$ and $\log N$, where $N$ is the largest coefficient appearing in the system \cite{Chou+SICOMP82}.
Then a solution $Y\in N(A)$ of the equation $X\ast Y\ast U_1=U_2\ast X\ast Y$ can be computed in time
polynomial in $\log\abs{A}$. The output of the algorithm is the matrix $X\ast Y$.
\end{proof}

\section{Our Algorithm}\label{section_algorithm}

In this section we give a proof of Theorem \ref{theorem_main}. 
We first present the following rather simple result that shows necessary and sufficient conditions
for the isomorphism of two groups in $\Sp$. 
\begin{proposition}\label{proposition_class}
Let $G$ and $H$ be two groups in $\Sp$.
Let $(A_1,\gen{y_1})$ and $(A_2,\gen{y_2})$ be standard decompositions
of $G$ and $H$ respectively and let $\varphi_1$ (resp. $\varphi_2$) be 
the action by conjugation of $y_1$ on $A_1$ (resp. of $y_2$ on $A_2$).
The groups $G$ and $H$ are isomorphic if and only if the following three conditions 
hold:
(i)
$A_1\cong A_2$; and 
(ii)
$\abs{y_1}=\abs{y_2}$; and
(iii)
there exists an integer $k\bound{1}{\abs{y_1}}$ coprime with $\abs{y_1}$ and an isomorphism $\psi:A_1\to A_2$
such that $\varphi_1=\psi^{-1}\varphi_2^k\psi$.
\end{proposition}
\begin{proof}
First notice that for a group $G$ in $\Sp$, the integer $\gamma(G)$  is a group invariant. 
Now suppose that $G$ and $H$ are two isomorphic groups in $\Sp$ with standard decomposition 
respectively $(A_1,\gen{y_1})$ and $(A_2,\gen{y_2})$. 
Then $\abs{y_1}=\abs{y_2}=\gamma(G)=\gamma(H)$.
Denote by $\psi$ an isomorphism from $G$ to $H$ and notice that $(\psi(A_1),\psi(y_1))\in\Dd^{\gamma(H)}_H$. 
From Lemma \ref{lemma_Hall} this implies that $\psi(A_1)=A_2$ and, in particular,  $A_1\cong A_2$. 
The element  $\psi(y_1)$ can be written as $zy_2^k$ for some $z\in A_2$ and some integer $k\bound{1}{\gamma(H)}$
coprime with $\gamma(H)$.
By definition of $\varphi_1$, for any $x\in A_1$ the relation $y_1x=(y_1xy_1^{-1})y_1=\varphi_1(x)y_1$ holds. 
Applying $\psi$ to each term gives 
\begin{eqnarray*}
zy_2^k\psi(x)&=&\psi(\varphi_1(x))zy_2^k\\
\varphi_2^k(\psi(x))zy_2^k&=&\psi(\varphi_1(x))zy_2^k
\end{eqnarray*}
for any $x\in A_1$. Thus $\varphi_2^{k}=\psi\varphi_1\psi^{-1}$.

Now consider two groups $G$ and $H$ in $\Sp$ satisfying the conditions (i), (ii) and (iii) of the statement of the theorem.
Denote $m=\abs{y_1}=\abs{y_2}$.
Let $\mu$ be the map from $G$ to $H$ such that $\mu(xy_1^j)=\psi(x_1)y_2^{kj}$ for any $x$ in $A_1$ and any 
$j\bound{0}{m-1}$. The map $\mu$ is clearly a bijection from $G$ to $H$. We now show that $\mu$ is a homomorphism,
and thus an isomorphism from $G$ to $H$. Let $x$ and $x'$ be two elements of $A_1$ and let $j$ and $j'$ be two elements
of  $\bound{0}{m-1}$. Then 
$$\mu(xy_1^jx'y_1^{j'})=\mu(x\varphi_1^j(x')y_1^{j+j'})=\psi(x\varphi_1^j(x'))y_2^{k(j+j')}=\psi(x)\psi(\varphi_1^j(x'))y_2^{k(j+j')}.$$
Now the relation
$\mu(xy_1^j)\mu(x'y_1^{j'})=\psi(x)y_2^{kj}\psi(x')y_2^{kj'}=\psi(x)\varphi_2^{kj}(\psi(x'))y_2^{k(j+j')}$ holds.
Condition (iii) of the statement of the theorem implies that $\psi(\varphi_1^j(x'))=\varphi_2^{kj}(\psi(x'))$ and thus
$\mu(xy_1^jx'y_1^{j'})=\mu(xy_1^j)\mu(x'y_1^{j'})$.
\end{proof}
{\bf Remark 1.}
Notice that the integer $k$ in Proposition \ref{proposition_class} cannot always be taken equal to $1$.
For example consider the groups $\gen{x_1,y_1\st x_1^7=y_1^3=e, y_1x_1=x_1^{2}y_1}$
and $\gen{x_2,y_2\st x_2^7=y_2^3=e, y_2x_2=x_2^{-3}y_2}$: 
the map $y_1\mapsto y_2^2$ and $x_1\mapsto x_2$ extends to an isomorphism
(because $y_2^2x_2=x_2^2y_2^2$) but no isomorphism mapping $y_1$ to $y_2$ exists.\\
\indent {\bf Remark 2.} Proposition \ref{proposition_class} can be used to give a (partial) mathematical classification of the 
number of groups of the form $A\rtimes \Int_m$. We refer to Appendix A for a sketch of how this can be done.\\

We now present our proof of Theorem \ref{theorem_main}.
\begin{proof}[Proof of Theorem \ref{theorem_main}]
Suppose that $G$ and $H$ are two groups in the class $\Sp$.
Denote $n=min(\abs{G},\abs{H})$ and 
$\gamma=min(\gamma(G),\gamma(H))$. 
In order to test whether these two groups are isomorphic, we first run
the algorithm of 
Theorem \ref{theorem_basis} on the inputs 
$G$ and $H$ and obtain outputs $(S_1,y_1)$ 
and $(S_2,y_2)$ such that $(\gen{S_1},\gen{y_1})$ and 
$(\gen{S_2},\gen{y_2})$ are standard decompositions of $G$ and $H$ 
respectively\footnote{Actually in order to obtain a running time bounded by $n$, and not by $max(\abs{G},\abs{H})$, 
we need to run the algorithm of Theorem \ref{theorem_basis} on the two inputs in parallel, compute the order of the group for which the algorithm first ends,
and stop the computation if the algorithm takes too long on the second input.}. 
The running time of this algorithm is $O(n^{1/2+o(1)})$
by Theorem \ref{theorem_basis}. Denote $A_1=\gen{S_1}$ and $A_2=\gen{S_2}$.

We then check whether $\abs{y_1}=\abs{y_2}$. If $\abs{y_1}\neq\abs{y_2}$ we conclude
that $G$ and $H$ are not isomorphic by Proposition \ref{proposition_class}. Otherwise notice that
 $\abs{y_1}=\abs{y_2}=\gamma$.
Then we compute a basis $(g_1,\ldots,g_s)$ of $A_1$ and a basis $(h_1,\ldots,h_t)$ of $A_2$ 
using the algorithm 
by Buchmann and Schmidt \cite{Buchmann+05}.
The running time of this step is $\tilde O(n^{1/2})$. 
Given these bases it is easy to check the isomorphism of $A_1$ and $A_2$:
the groups $A_1$ and $A_2$ are isomorphic if and only if $s=t$ and $\abs{g_i}=\abs{h_i}$
for each $i\bound{1}{s}$.
If $A_1\not\cong A_2$ we conclude that $G$ and $H$ are not 
isomorphic by Proposition \ref{proposition_class}.

Now suppose that $A_1\cong A_2$ (and then $s=t$) and denote $R=R(A_1)=R(A_2)$. 
We want to decide whether 
the action by conjugation $\varphi_1$ of $y_1$ on $A_1$ 
and the action by conjugation $\varphi_2$ of $y_2$ on $A_2$ 
satisfy Condition (iii) in Proposition \ref{proposition_class}.
Let $p_1^{d_1}\cdots p_r^{d_r}$ be the prime power decomposition of $\abs{A_1}=\abs{A_2}$, 
with $p_1<\cdots<p_d$ and denote $P_i$ the Sylow $p_i$-subgroup of $A_1$ for each $i\bound{1}{r}$. 
%
We compute the matrix $M_1$ in $R$ corresponding to the automorphism $\varphi_1$ of $A_1$
with respect to the basis $(g_1,\ldots,g_s)$. More precisely let us denote
$\varphi_1(g_i)=y_1g_iy_1^{-1}=g_1^{u_{i1}}\cdots g_j^{u_{is}}$ for each $i\bound{1}{s}$. The values 
$u_{ij}$ for each $i$ can be found by using the algorithm of Proposition \ref{proposition_abmem} 
on the input $y_1g_iy_1^{-1}$. Then the matrix $M_1=(u_{ij})$ can be computed in
time $\tilde O(n^{1/2})$. Similarly we compute the matrix $M_2\in R$ corresponding to the automorphism 
$\varphi_2$ of $A_2$ with respect to the basis $(h_1,\ldots,h_s)$. 
 A key observation is that $M_1$ and $M_2$ are block diagonal, consisting in $r$ blocks. More precisely
 the $i$-th block is a matrix in $R(P_i)$.

Finally for each integer $k\bound{1}{\gamma}$ coprime with $\gamma$, we test whether $M_1$ and $M_2^k$ are conjugate in $R$.
This is done by using the algorithm of Theorem \ref{theorem_conj} to check whether, for each $i\bound{1}{r}$, 
the $i$-th block of $M_1$ is conjugate to the $i$-th block of $M_2$ in $R(P_i)$.
If there is no $k$ such that $M_1$ and $M_2^k$ are conjugate in $R$ we conclude that $G$ and
$H$ are not isomorphic. Otherwise we take one value $k$ such that $M_1$ and $M_2^k$ are conjugate and compute
an explicit block diagonal matrix $X$ in $R$ such that $M_1=X^{-1}M_2^kX$. 
This can be done in time polynomial in $\log n$ by Theorem \ref{theorem_conj}. 
The matrix $X$ is naturally associated to an isomorphism $\psi$ from $A_1$ to $A_2$ through the bases
$(g_1,\ldots,g_s)$ and $(h_1,\ldots,h_s)$.
The map $\mu: G\to H$ defined as $\mu(xy_1^{j})=\psi(x)y_2^{kj}$ for any $x\in A_1$ and any $j\bound{0}{\gamma-1}$ is 
then an isomorphism from $G$ to $H$ (see the proof of Proposition \ref{proposition_class} for details).
The total complexity of this final step is  $O(\gamma\log^{c}n)$ for some constant $c$. 

The global time complexity of this algorithm is $O(\gamma\log^{c}n)+O(n^{1/2+o(1)})\le (\sqrt{n}+\gamma)^{1+o(1)}$.
\end{proof}

\section*{Acknowledgments}
The author is grateful to Yoshifumi Inui for many discussions on similar topics.
He also thanks Igor Shparlinski and Erich Kaltofen for helpful comments.

\newpage
\noindent {\LARGE \bf Appendix}\vspace{3mm}\\
\noindent {\Large \bf A The Number of Isomorphism Classes}\vspace{4mm}

We briefly mention a sketch of how our results can be used to derive the number of isomorphism
classes of groups of the form $A\rtimes \Int_m$ for some given abelian group $A$ and positive integer $m$ such that
$gcd(\abs{A},m)=1$.  We only work out the rather simple case where $A=\Int_{3^i}^r$ and $m=4$ here.
We believe that this gives an insight of the usefulness of our results and of the rich mathematical structure of 
the class of groups $\Sp$.

Let $G=A\rtimes \gen{y}$ where $\abs{y}=4$ and $A=\Int_{3^i}^r$ for some positive integers $i$ and $r$.
Then the action by conjugation of $y$ over $A$ can be written as a matrix $M$ in $R(A)$. Notice that necessarily
$M^4=I$.
From proposition \ref{proposition_class} two distinct actions $M_1$ and $M_2$ define isomorphic
groups if and only $M_1$ and $M_2^k$ are conjugate in $R(A)$ for $k=1$ or $k=3$. 
We will show that, for $A=\Int_{3^i}^r$ and $m=4$,  the matrices $M_1$ and $M_2$ are conjugate in $R(A)$
if and only if  $M_1$ and $M_2^3$ are conjugate in $R(A)$. Thus, in this case, Propositions 
\ref{proposition_conj} and \ref{proposition_class}  imply that  $M_1$ and $M_2$
define isomorphic groups if and only if $\Psi(M_1)$ and $\Psi(M_2)$ are conjugate
in $V(A)=GL_r(3)$. We stress that for other values of $A$ and $m$ this is not always the case (see for example the isomorphic groups
$\gen{x_1,y_1\st x_1^7=y_1^3=e, y_1x_1=x_1^{2}y_1}$
and $\gen{x_2,y_2\st x_2^7=y_2^3=e, y_2x_2=x_2^{-3}y_2}$ already mentioned in Section \ref{section_algorithm}).

The number of conjugacy classes of matrices of a given order $s$ in the general linear
group $GL_r(p)$ is well known \cite{Hodges58}, although usually difficult to write down in a concise way.
This number is related to the factorization of the polynomial $X^s-1$
in the field $\Int_p$ through the concept of the canonical normal form of a matrix. For example for the values $s=4$ and $p=3$ 
the factorization is $X^4-1=(X+1)(X-1)(X^2+1)$. 
Let $U$, $V$ and $W$ be the companion matrices associated to the polynomial $X+1$, $X-1$, and $X^2+1$ 
respectively. Then properties of the canonical normal form show that any matrix of order dividing $4$ in $GL_r(3)$ is conjugate to a unique block diagonal matrix 
where the first $k_1$ blocks are $U$, the next $k_2$ blocks are $V$ and the last $k_3$ blocks are $W$,
for some $(k_1,k_2,k_3)\in S_r$. Here $S_r$ denotes the set $\{(k_1,k_2,k_3)\in\Int^3\st k_1\ge 0,\: k_2\ge 0,\: k_3\ge 0, \: k_1+k_2+2k_3=r\}$. 
Thus there are $\abs{S_r}$ conjugacy classes of matrices of order dividing 4 in $GL_r(3)$.

A key observation is now that $U^3$ is conjugate to $U$ in $GL_1(3)$, 
$V^3$ is conjugate to $V$ in $GL_1(3)$ and $W^3$ is conjugate to $W$ in $GL_2(3)$. Thus 
if $M_1$ and $M_2^3$ are conjugate in $R(A)$ then necessarily $M_1$ and $M_2$ 
are conjugate in $R(A)$ too. We conclude that the number of isomorphism types for the groups 
$\Int_{3^i}^r\rtimes\Int_4$ is $\abs{S_r}$. For example the number of isomorphism types for 
$\Int_3^4\rtimes\Int_4$ is 9, as mentioned in the introduction of this paper.


\vspace{4mm}

\end{document}